\documentclass[aps,prb,superscriptaddress,twocolumn]{revtex4}
\usepackage{graphicx}
\usepackage{amsmath}
\usepackage{bm}
\usepackage{xcolor}

\begin{document}

\title{Transport Induced Dimer State from Topological Corner States}

\author{Kai-Tong Wang}
\affiliation{College of Physics and Energy, Shenzhen University, Shenzhen 518060, China}
\affiliation{Key Laboratory of Optoelectronic Devices and Systems of Ministry of Education and Guangdong Province, College of Optoelectronic Engineering, Shenzhen University, Shenzhen 518060, China}

\author{Yafei Ren}
\affiliation{Department of Physics, The University of Texas at Austin, Austin, Texas 78712, USA}

\author{Fuming Xu}
\email[]{xufuming@szu.edu.cn}
\affiliation{College of Physics and Optoelectronic Engineering, Shenzhen University, Shenzhen 518060, China}

\author{Yadong Wei}
\affiliation{College of Physics and Optoelectronic Engineering, Shenzhen University, Shenzhen 518060, China}

\author{Jian Wang}
\email[]{jianwang@hku.hk}
\affiliation{College of Physics and Optoelectronic Engineering, Shenzhen University, Shenzhen 518060, China}
\affiliation{Department of Physics, University of Hong Kong, Pokfulam Road, Hong Kong}
\begin{abstract}
Recently, a new type of second-order topological insulator has been theoretically proposed by introducing an in-plane Zeeman field into the Kane-Mele model in the two-dimensional honeycomb lattice. A pair of topological corner states arise at the corners with obtuse angles of an isolated diamond-shaped flake. To probe the corner states, we study their transport properties by attaching two leads to the system. Dressed by incoming electrons, the dynamic corner state is very different from its static counterpart. Resonant tunneling through the dressed corner state can occur by tuning the in-plane Zeeman field. At the resonance, the pair of spatially well separated
and highly localized corner states can form a dimer state, whose wavefunction extends almost the entire bulk of the diamond-shaped flake. By varying the Zeeman field strength, multiple resonant tunneling events are mediated by the same dimer state. This re-entrance effect can be understood by a simple model. These findings extend our understanding of dynamic aspects of the second-order topological corner states.
\end{abstract}

\maketitle

\section{Introduction}
Two-dimensional insulators with nontrivial band topology have been widely studied in the past decades~\cite{Qi2010,Hasan2010,Ren2015,Bansil2016}, which include quantum anomalous Hall effect~\cite{1988Model,Qi2005,Yu2010,Qiao2010,Chang2016,Ren2017}, quantum spin Hall effect~\cite{Kane2005,Bernevig2006,Roth2009,Qi2016Effective}, quantum valley Hall effect~\cite{Xiao2007,Martin2007,Qiao2011,Anglin2016,Li2017,Hou2018,Cheng2018,Qiao2013,Jung2011,Ying2013,Bi2015,Lee2016,
2016Valley,2015Topological,Zhang2013,2019On,2016Gate}, and topological crystalline insulators~\cite{Fu2010,Ando2015,Chiu2015}. These topological phases are characterized by nontrivial band topological indices as well as gapless edge states, which are robust against disorders and exhibit quantized conductivity~\cite{1988Model,Qi2005,Yu2010,Qiao2010,Chang2016,Roth2009,Xing2011,2014Universal,He2013,2015Quantum,Weng2015,Zhang2016The}.

Recently, higher-order topological insulators (TIs) have been theoretically proposed in various systems~\cite{Slager2015,Benalcazar2016,Benalcazar2017,Khalaf2018,SerraGarcia2018,Klinovaja2019,Chen2019,Hu2019,Huang2018,Schindler2019}, which are characterized by hinge modes in three-dimensional (3D) materials~\cite{Langbehn2017,Song2017,Miert2018,Schindler2017} or corner states in two-dimensional (2D) systems~\cite{Ezawa2018,Ezawa2017,Sheng2019,Benalcazar2018,Park2019,Zeng2019,Li2019}. The existence of one-dimensional hinge states has been experimentally confirmed in bismuth~\cite{Schindler2018} and multi-layer WTe$_2$~\cite{Choi2019}. In 2D higher-order TIs, one-dimensional edges are insulating whereas the corners between different edges can host zero-dimensional in-gap states that are isolated from both edge and bulk bands by an energy gap~\cite{Ezawa2018,2019Two,Yang2020Gaped,2020Two,Sheng2019,Park2019,Chen2020}. In contrast to the gapless edge states in conventional topological insulators, higher-order topological corner states are not conducting and behave like localized bound states. Therefore, it is challenging to detect these corner states by transport measurement.

In this work, we numerically study the transport properties of topological corner states in the two-dimensional honeycomb lattice. Based on the modified Kane-Mele model with an in-plane Zeeman field, the second-order topological insulator is realized and zero-energy corner states are localized at the intersections of different zigzag boundaries of an isolated diamond-shaped flake~\cite{Ren2020}. By connecting two leads to the diamond-shaped flake while keeping the corner states intact, dynamic features of corner states are revealed. Different from the static corner state, incident electrons dwelling in the corner state can cause its "delocalization" that mediates the transport. By tuning the Zeeman field, corner-state-mediated resonant tunneling occurs. For a single corner-state setup, as the Zeeman field is increased, resonant tunneling via the corner state or its precursor persists, until a threshold field strength closes the resonant channel. For a setup with two corner states, a dimer state is formed at the resonance with its wavefunction spanning almost the entire bulk of the flake. As the Zeeman field is scanned, it is found that after the corner states emerge, multiple resonant peaks mediated by the dimer state can arise in the transmission spectrum, which is counter-intuitive since there is only one dimer state. A simple model is constructed, which explains that this re-entrance effect is indeed caused by the same dimer state.

\section{Model and Formalism}

In the conventional Kane-Mele model\cite{Kane2005}, intrinsic spin-orbit coupling gives rise to quantum spin Hall effect in the two-dimensional honeycomb lattice. By applying an in-plane Zeeman field, the time-reversal symmetry is broken and higher-order topological states emerge~\cite{Ren2020}. This modified Kane-Mele model has the following tight-binding Hamiltonian~\cite{Ren2020}:
\begin{align}
H_D = -t\sum_{<ij>} c_{i}^{\dag}c_{j} + it_{so}\sum_{\ll ij \gg} \nu_{ij}c_{i}^{\dag}s_{z}c_{j}
       + \lambda \sum_{i} c_{i}^{\dag} \mathbf {B}\cdot \mathbf {s}c_{i},\label{Eq1}
\end{align}
where $c_{i}^{\dag} = [c_{i,\uparrow}^{\dag},c_{i,\downarrow}^{\dag}]$ is the creation operator at site $i$ with spin up and spin down. $t$ is the nearest-neighbor hopping energy. The second term represents the intrinsic spin-orbit coupling with $t_{so}$ the coupling strength, which only involves the next-nearest-neighbor hopping and $\nu_{ij}= +1/-1$ if the electron moves from site $j$ to site $i$ by taking a left/right turn. The last term originates from the in-plane Zeeman field, whose direction depends on the magnetic field. $\mathbf {s}$ is the Pauli matrix for spin and $\lambda$ denotes the Zeeman field strength. We set $t_{so}=0.1t$ and choose the magnetic field along $y$-direction as $\mathbf{B}=(0,B_y,0)$. For convenience, $t$ is set as the energy unit.

\begin{figure}
\includegraphics[width=8.5cm]{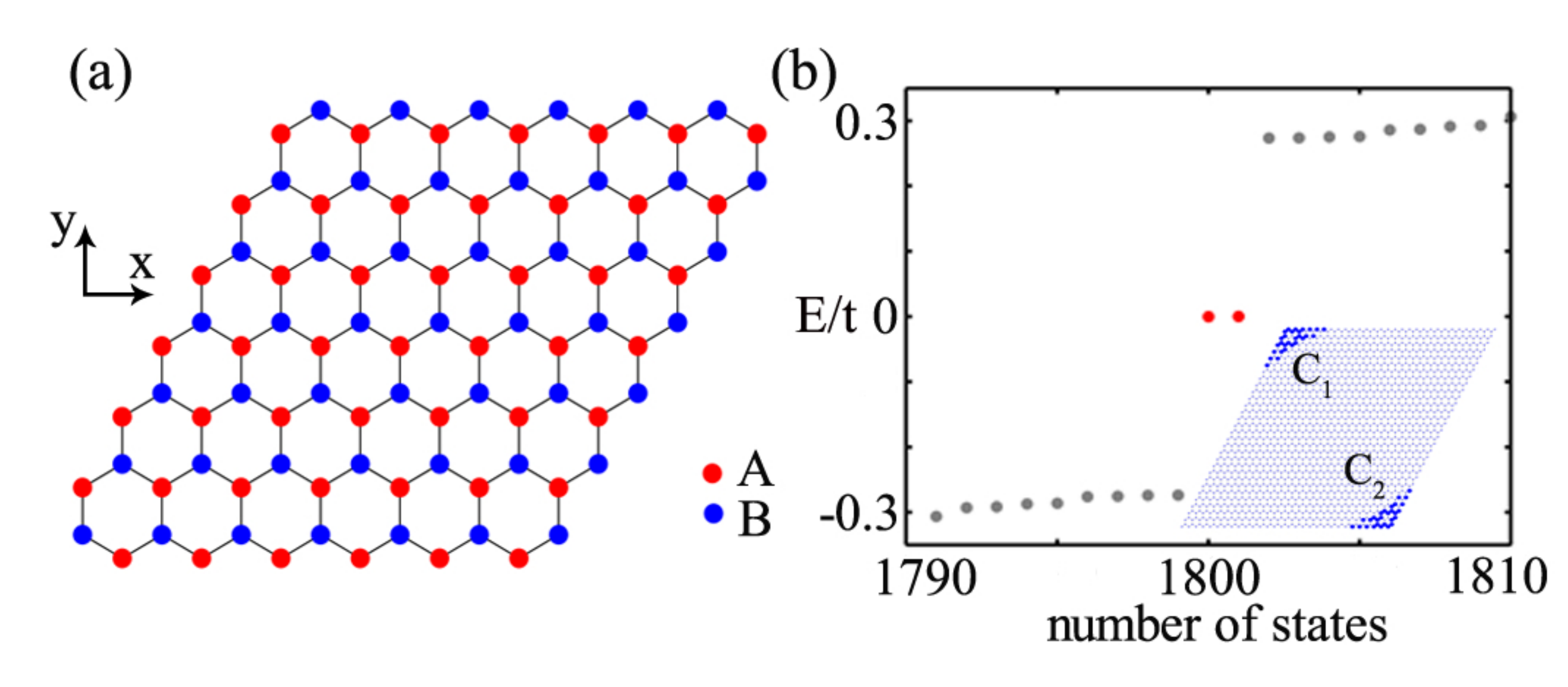}
\caption{(Color online) (a) Sketch of a diamond-shaped honeycomb-lattice flake with zigzag boundaries. Red and blue dots denote A/B sublattices, respectively. (b) Energy levels for the diamond-shaped flake with $\lambda=0.3$. Eigenfunction distributions of the corner states are shown in the inset, and the corresponding eigenenergies are highlighted in red. $C_1$ and $C_2$ are the intersections of zigzag boundaries where corner states emerge. }
\label{fig1}
\end{figure}

We consider the diamond-shaped honeycomb-lattice flake surrounded by zigzag boundaries as shown in Fig.\ref{fig1}(a). In the absence of the Zeeman field, gapless edge states distribute continuously along zigzag boundaries of the flake. When the Zeeman field is turned on, the edge states start to shrink towards the corners with obtuse angles while delocalizing away from the flake boundaries. Meanwhile, the zero-energy levels stay unchanged in the energy spectrum while non-zero energy levels are repelled from $E=0$. As the Zeeman field is increased to $\lambda \ge 0.2$~\cite{Ren2020}, a large energy gap opens up accompanied by a pair of zero-energy states, which are highlighted in red in Fig.\ref{fig1}(b). The corresponding eigenfunctions of these zero-energy states are localized around $C_1$ and $C_2$ corners of the diamond-shaped flake as illustrated in the inset. These zero-energy corner states manifest the second-order topological phase characterized by nonzero winding numbers of the bulk band~\cite{Ren2020}.

To investigate transport behaviors of the corner states, the diamond-shaped flake is connected to two conducting leads to calculate the conductance or transmission of the system. The full Hamiltonian of the transport system is given by
\begin{align}
H = H_L + H_D + H_T,
\end{align}
where $H_L$ is the lead Hamiltonian following the Kane-Mele model\cite{Kane2005}
\begin{align}
H_L = -t\sum_{<nm>} d_{n}^{\dag}d_{m} + it_{so}\sum_{\ll nm \gg} \nu_{nm}d_{n}^{\dag}s_{z}d_{m},
\end{align}
with $d_{n}^{\dag}$ representing the creation operator in the lead at site $n$. The second term $H_D$ is defined in Eq.(\ref{Eq1}), which is the Hamiltonian of the central diamond-shaped region. The third term $H_T$ is the Hamiltonian describing the coupling between the central region and the leads
\begin{align}
H_T = -t\sum_{<nj>} d_{n}^{\dag}c_{j} + it_{so}\sum_{\ll nj \gg} \nu_{nj}d_{n}^{\dag}s_{z}c_{j}+ H.c. .
\end{align}
Following the Landauer-B$\rm\ddot{{u}}$ttiker formula, the transmission from lead $q$ to lead $p$ is expressed as
\begin{equation}\label{Trans}
T_{pq} =  {\rm Tr}[\Gamma_{q} G^{r} \Gamma_{p} G^{a}].
\end{equation}
Here $G^{r}$ and $G^{a}$ are the retarded and advanced Green's functions of the central scattering region, which are defined as $G^{r}(E) = (E-H_D-{\Sigma}^{r})^{-1}$ and $G^{a}={G^{r}}^\dag$. ${\Sigma}^r={\Sigma}^{r}_{p}+{\Sigma}^{r}_{q}$ is the self-energy contributed by both leads, which are iteratively constructed through the transfer-matrix method.\cite{Lopez1984jpf} $\Gamma_{p/q} =i({\Sigma}_{p/q}^{r}-{\Sigma}_{p/q}^{a})$ is the linewidth function representing the interaction between leads and the central region.

The nature of corner states can be examined both statically and dynamically: (1) for an isolated system, the eigenfunction distribution gives the static picture of corner states; (2) for an open system, the partial local density of states (LDOS) dynamically describes how an electron incident from lead $p$ dwells inside the scattering region. The partial LDOS is defined as
\begin{equation}\label{Ldos}
\rho_p(j) = \frac{1}{2\pi} [G^{r} \Gamma_{p} G^{a}]_{jj},
\end{equation}
with $j$ denoting the real-space site. The calculation of partial LDOS can be effectively accelerated by using the eigenstate form of $\Gamma_{p}$~\cite{Wang2009prb,Xu2011prb}.

\section{Numerical Results and Discussion}

Several transport setups are proposed to probe the localized corner states and reveal their transport characteristics, by connecting the diamond-shaped flake to zigzag honeycomb-lattice leads in different ways. The size of this flake is fixed as $L_{x}\times L_{y}=30a\times 30a$ with $a$ being the lattice constant. Unless specified otherwise, eigenfunction for the isolated system, partial LDOS and transmission for the open system, are all evaluated at zero energy. Four possible setups are investigated. Our numerical results show that, corner states are destroyed once the leads are directly connected to $C_1$ and $C_2$ regions of the flake. Two setups shown in Fig.\ref{figs1}(a) and Fig.\ref{figs2}(a) belong to this category, where detailed results and analysis are presented in the supplemental material. In the following, we consider the other two setups. In the first setup, only one corner state survives, which can be probed by resonant tunneling. In the second setup, the leads are far away from $C_1$ and $C_2$ so that both corner states remain intact. In this case, a counter-intuitive resonant tunneling phenomenon mediated by both corner states occurs. During this tunneling process, the incident electron traverses almost the entire region of the flake, which is the classically forbidden region. The dressing of incoming electrons leads to the formation of a dimer state, which is symmetrically distributed among both corner states.

\subsection{Single Corner-State Setup}

\begin{figure}
\includegraphics[width=8.5cm,clip=]{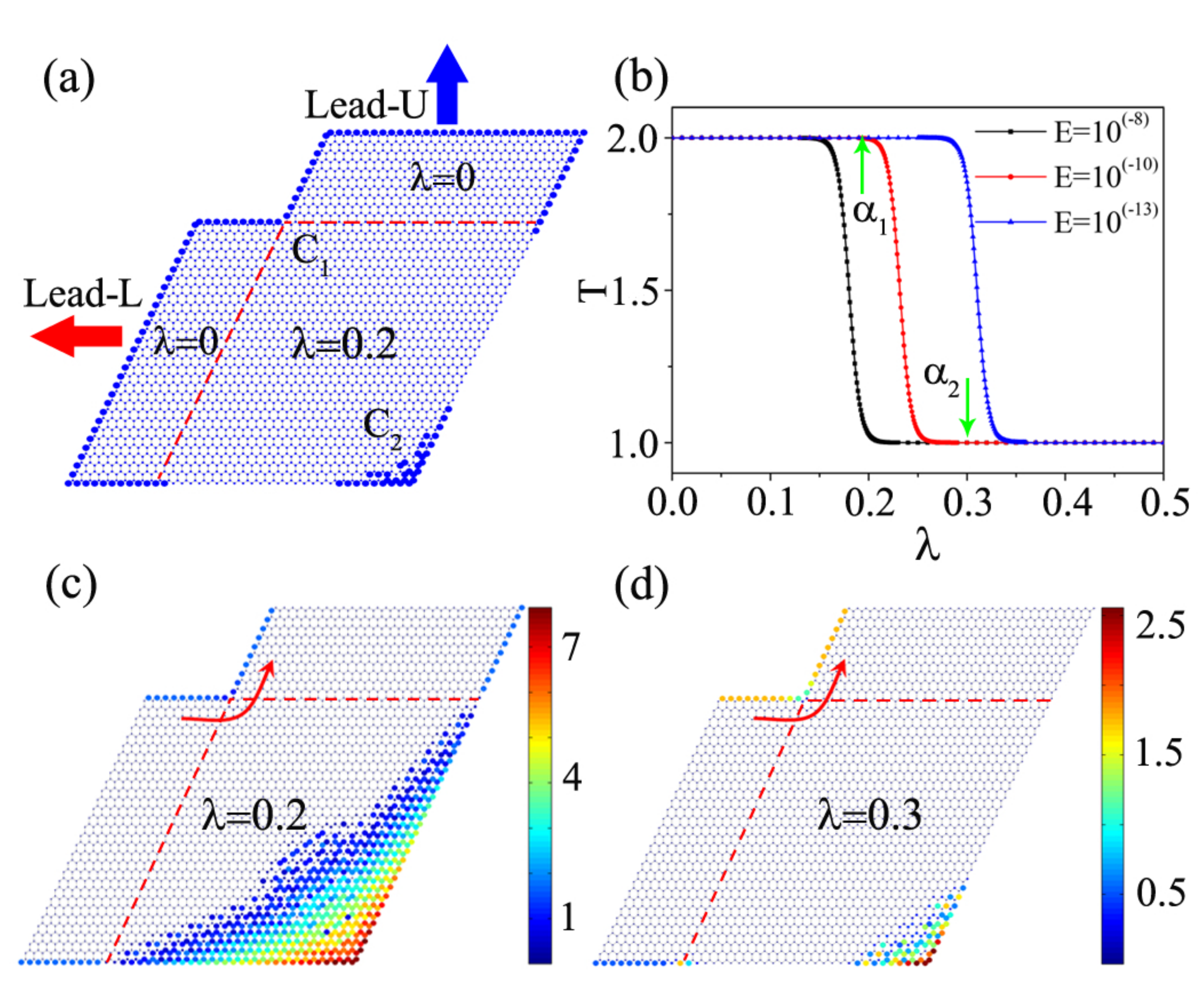}
\caption{ (Color online) (a) Schematic of the single corner-state setup coupled with left and top leads, together with the eigenfunction distribution shown in blue. (b) Transmission versus the Zeeman field strength for different energies $E$. (c) Partial LDOS for $E=10^{-13}$, i.e., the $\alpha_1$ ($\lambda=0.2$) point for $T$=2 in panel (b). (d) Partial LDOS for $E=10^{-8}$, which is the $\alpha_2$ point for $T$=1. }\label{fig2}
\end{figure}

In the single corner-state setup shown in Fig.\ref{fig2}(a), left and top leads are attached to the central region. Edge states along zigzag boundaries of the $\lambda$=0 regions are shown in blue. Single corner state arises at the $C_2$ corner with an exponentially decaying wavefunction. The dependence of the transmission $T$ on the Zeeman field strength for different energies are plotted in Fig.\ref{fig2}(b) showing two transmission plateaus with $T$=2 and $T$=1. For a particular incident energy, there is a critical strength of the Zeeman field beyond which the transmission drops. When the incident energy is further away from zero, the $T=2$ plateau is more easily destroyed by the Zeeman field. These transmission plateaus usually manifest the signature of quantized transport. However, the $T$=2 transmission plateau is contributed from two distinct conducting channels, one from the edge-state channel and the other from the corner-state-mediated resonant tunneling channel. To understand the physical origin of transmission plateaus, we plot the partial LDOS for open systems in Fig.\ref{fig2}(c) and Fig.\ref{fig2}(d). Given $\lambda$=0.2 and $E=10^{-13}$ for $T=2$, the incoming electron tunnels through the flake via the corner state at $C_2$. Due to the dressing of incoming electrons, the corner state becomes delocalized and its wavefunction expands into the classically forbidden bulk region. Two conducting channels are clearly seen in Fig.\ref{fig2}(c): one connects edge states along zigzag boundaries of both leads as labeled by the red arrow near $C_1$, the other is formed through the corner state at $C_2$. When the Zeeman field is increased further, the corner-state wavefunction shrinks towards the $C_2$ corner until it is completely localized and the corresponding conducting channel is closed. Thus, only one edge-state channel survives in Fig.\ref{fig2}(d) resulting in the $T$=1 plateau.

Now we study the typical resonant feature. The transmission vs the incident energy is shown in Fig.S3(a) of the supplemental material, which exhibits extremely sharp peaks near $E=0$. The $T=2$ transmission peak is easily destroyed by either changing the incident energy or increasing the Zeeman field strength, showing its resonant nature. Meanwhile, the $T=1$ plateau is robust against both $\lambda$ and $E$, since it originates from the edge state connecting left and top leads at the $C_1$ corner in Fig.\ref{fig2}(c). To further confirm the resonant nature, we define $\Delta E$ as the half-width energy at half-maximum of the resonant peak ($T=1.5$). Apparently, the stronger the Zeeman field, the smaller the $\Delta E$. A stronger Zeeman field leads to more localized corner states, and it is easier to close the resonant channel through the $C_2$ corner. $\Delta E$ can also be obtained by solving the eigenvalue problem of the effective Hamiltonian including the effect of both leads. The imaginary part of the eigenvalue is identified as the lifetime of the resonant state, which is equivalent to $\Delta E$. Numerical results presented in the supplemental material show that this is indeed the case.

\subsection{Double Corner-State Setup}

\begin{figure}
\includegraphics[width=8cm ]{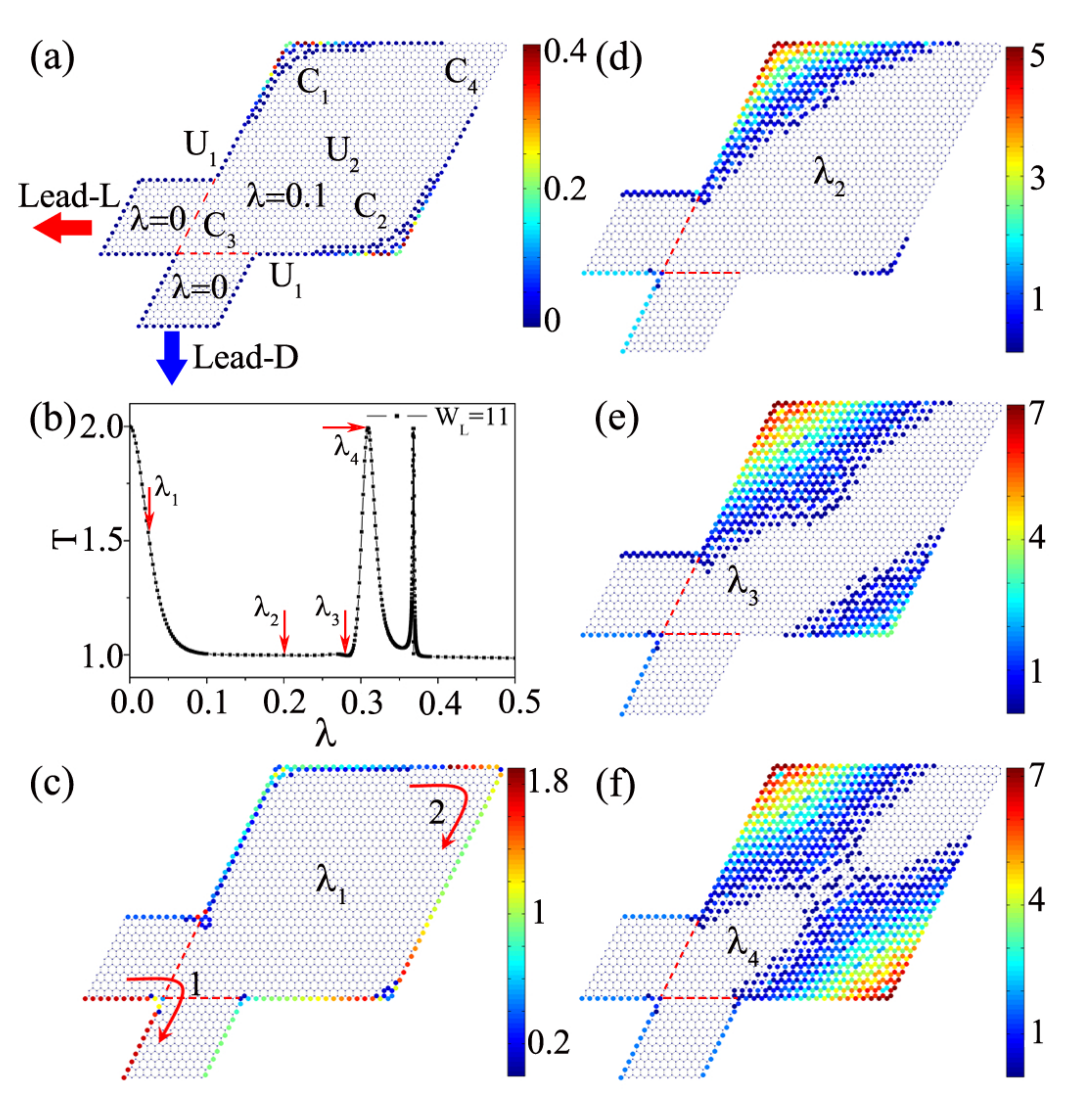}
\caption{ (Color online) (a) The double corner-state setup, where the central region is connected to narrower left and bottom leads around the $C_3$ corner. (b) Transmission with lead width $W_L$=11. Four typical $\lambda$ are shown in red arrows, which are $0.024$, $0.2$, $0.28$, and $0.31$, respectively. Panels (c)-(f): Partial LDOS for $\lambda_1$-$\lambda_4$.}\label{fig3}
\end{figure}

As shown in Fig.\ref{fig3}(a), the diamond-shaped flake is connected to narrower leads around the $C_3$ corner. Both leads are far away from $C_1$ and $C_2$ to ensure weak impact on the corner states. The zero-energy eigenfunction of this isolated system with $\lambda=0.1$ shows the precursor of two corner states at both $C_1$ and $C_2$. Transmission as a function of the Zeeman field is displayed in Fig.\ref{fig3}(b). At $\lambda=0$, there are two conducting channels due to the edge states giving rise to $T=2$. When the Zeeman field is applied and increased, $T$ decreases from $T=2$ and quickly reaches the $T$=1 plateau, which indicates the closing of one conducting channel. Remarkably, with further increasing of $\lambda$, multiple resonant peaks of $T=2$ emerge. As discussed in the supplemental material, the second sharp peak is actually double peaks and the influence of lead width is also evaluated. In view of resonant transport, these integer transmission peaks of $T=2$ manifest the existence of two perfect propagating channels in the system. In the following, we will examine the nature of this resonant tunneling and explain why multiple resonant peaks appear.

We focus on the setup with lead width $W_L$=11 and plot partial LDOS of the open system in Fig.\ref{fig3}(c-f), which correspond to four typical Zeeman field strengths labeled as $\lambda_1$ to $\lambda_4$ in Fig.\ref{fig3}(b). When $\lambda$=0, two edge states of quantum spin-Hall nature dominate the transport, which leads to $T=2$. For a small $\lambda_1$=0.024, two conducting channels are still visible in Fig.\ref{fig3}(c). But the channel along boundary 2 is not transmissive enough so that it can only sustain a $T=1.5$ transmission. For a larger strength $\lambda_2$=0.2 at the $T$=1 plateau, only the channel along boundary 1 survives and the other channel is destroyed (Fig.\ref{fig3}(d)). Nevertheless, electrons incident from the left lead can still tunnel through the barrier $U_1$ labeled in Fig.\ref{fig3}(a) and reach the upper corner $C_1$ with large partial LDOS, which shows the dynamic signature of the dressed corner state. However, it can not reach the lower corner $C_2$ leaving the static corner state completely isolated. Fig.\ref{fig3}(d) shows the distinction between static and dynamic corner states located at $C_2$ and $C_1$, respectively. When $\lambda$ is increased to $\lambda_3$, Fig.\ref{fig3}(e) shows the dynamic feature of corner states at both $C_1$ and $C_2$, which expand and grow inside the bulk due to the dressing of incoming electrons. Counter-intuitive phenomenon happens at $\lambda_4$. In Fig.\ref{fig3}(f), dynamic corner states at $C_1$ and $C_2$ bridge the spatial gap inside the system and form a dimer state, which leads to a new resonant channel inside the bulk instead of along the boundaries of the flake. As a result, a resonant peak with $T$=2 appears, which is sharp and sensitive to both $\lambda$ and $E$. Numerical results verify that, all the resonant peaks in Fig.\ref{fig3}(b) originate from the same mechanism: the dimer-state-mediated resonant tunneling. The nature of the dimer state has been studied in detail in the supplemental material. To identify the dimer state, the following observations are in order: (1) the dressing of incoming electrons leads to the binding of two corner states; (2) the distribution of the dimer wavefunction should be symmetrical in order to mediate the resonant tunneling; (3) comparing Fig.\ref{fig3}(f) with Fig.S5(b) the binding of two corner states is closely related to the sharpness of the resonant peak.

\begin{figure}
\includegraphics[width=8.5cm ]{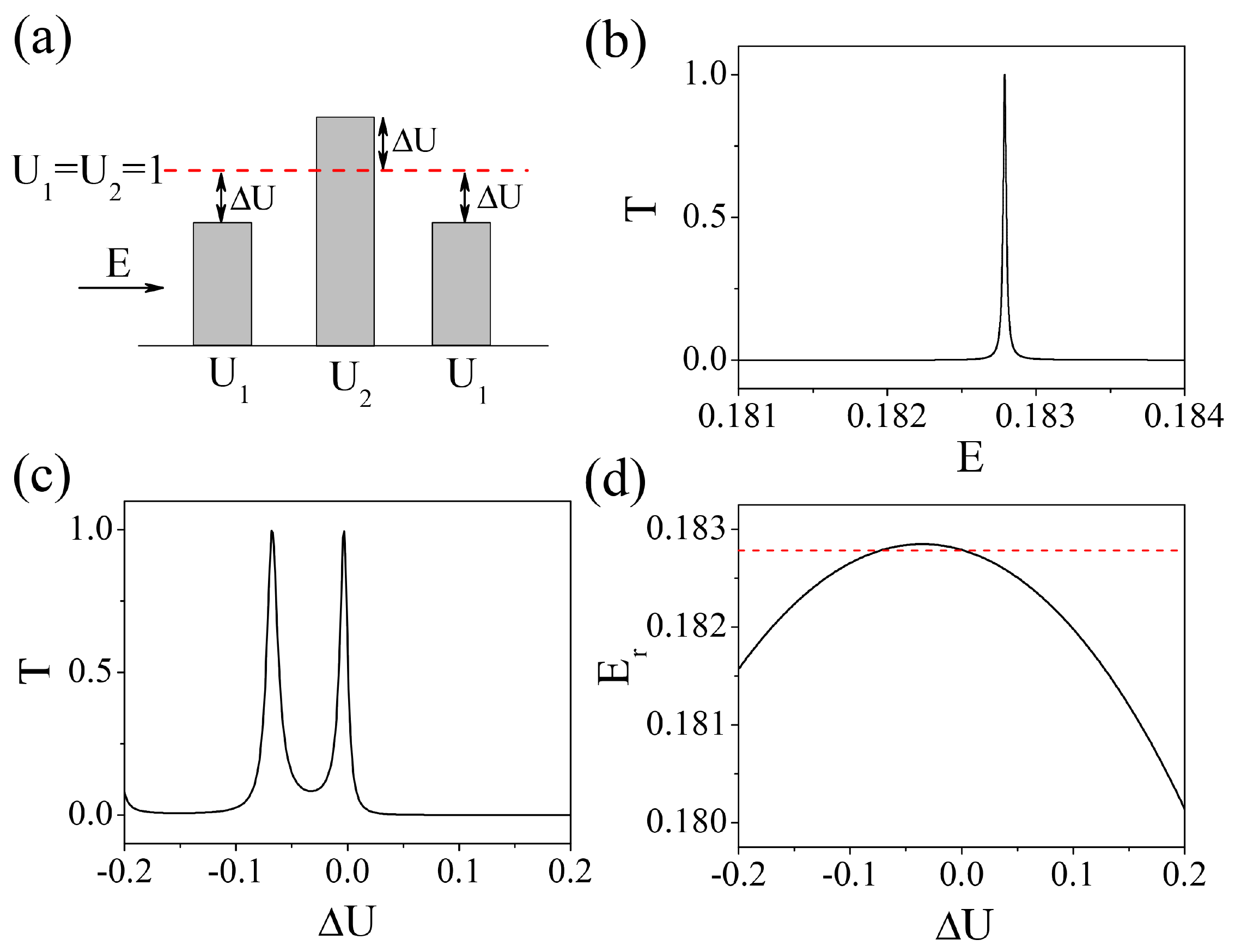}
\caption{(Color online) (a) The triple-barrier model in one-dimensional tight-binding chain~\cite{sdatta}. $U_{1/2}$ denotes the strength of potential barriers with $\Delta U$ the relative variation. (b) Transmission spectrum for $U_{1/2}=1$. (c) Transmission with respect to the relative variation $\Delta U$ for $U_1=1+\Delta U$ and $U_2=1-\Delta U$ at the resonant level $E_{res}=0.1828$. (d) Real part $E_r$ of the complex eigenenergies versus $\Delta U$. Red dash line corresponds to $E_{res}=0.1828$. }\label{fig4}
\end{figure}

The existence of multiple resonant peaks with respect to the Zeeman field strength, instead of a single one, can be understood with a simple model. We note that in the presence of the Zeeman field, there are three spatial barriers in Fig.\ref{fig3}(a) for an incident electron to overcome in resonant tunneling. $U_1$ is the effective potential between the leads and the corner states, and $U_2$ is that in the bulk separating $C_1$ and $C_2$. The evolution of $U_1$ and $U_2$ follows that of the corner state. As the Zeeman field is increased, the edge state shrinks along the boundary and delocalizes away from it so that the strength of $U_1$ increases while $U_2$ decreases. In another word, the Zeeman field strength controls the barrier variation in this system, which is confirmed by partial LDOS shown in Fig.\ref{fig3}. We construct a triple-barrier model based on a one-dimensional tight-binding chain~\cite{sdatta}, where barrier strengths $U_1$ and $U_2$ are simultaneously varied with a relative amount $\Delta U$ (see Fig.\ref{fig4}(a)). The tight-binding chain has $35$ lattice sites and all three barriers with width $5$ sites are evenly spaced. When $U_{1/2}=1$, the resonant level for this triple-barrier system is found to be $E_{res}$=0.1828 as shown in Fig.\ref{fig4}(b). By fixing $E_{res}$ and tuning the effective potentials with $U_1=1+\Delta U$ and $U_2=1-\Delta U$, the transmission with respect to $\Delta U$ shows two resonant peaks with $T=1$ as shown in Fig.\ref{fig4}(c). These peaks are similar to the multiple resonant peaks in Fig.\ref{fig3}(b). Here the relative variation $\Delta U$ plays a similar role as the Zeeman field strength $\lambda$ in Fig.\ref{fig3}(b). The evolution of the resonant level $E_{res}$ of the triple-barrier system is plotted in Fig.\ref{fig4}(d). The procedure of calculating $E_r$, which is the real part of the complex eigenenergy, is demonstrated in the supplemental material. We see that $E_r$ crosses the incident energy twice, which gives rise to two resonant peaks in Fig.\ref{fig4}(c). Hence these two resonant peaks are mediated by the same resonant state showing the re-entrance effect. Similarly, by varying the Zeeman field strength in the double corner-state setup, multiple resonant tunneling processes are mediated by the same dimer state.

\section{Conclusion}

In summary, by connecting two leads to the system hosting topological corner states, we studied dynamic features of the corner state. Dressed by incoming electrons, the localized static corner state becomes extended, which can give rise to corner-state-mediated resonant tunneling. For a setup with a single corner state, by tuning the Zeeman field, we observed resonant tunneling mediated by a state that evolves from the edge state, the precursor of a corner state, and the corner state, and finally the closing of this resonant channel with a large enough Zeeman field. When two corner states are present, the resonant tunneling can also occur. At the resonance, the dressing of incoming electrons can bind two corner states to form a dimer state, whose wavefunction extends to most of the bulk region that is classically forbidden. As the Zeeman field strength is varied, the dimer-state-mediated resonant tunneling exhibits re-entrance effect with multiple resonant peaks, which can be understood by a simple triple-barrier model. Given the advancement of modern technologies, our proposal for probing the corner state and its dynamic features can be realized experimentally.

This work was financially supported by the Natural Science Foundation of China (Grant No. 12034014), Guangdong Province (Grant No. 2020A1515011418), and Shenzhen (Grants No. JCYJ20190808152801642 and JCYJ20190808150409413).

\widetext

\section{Supplemental materials}

\subsection*{S1. Setup I: the central region coupled with left and right leads}

\begin{figure}[tbp]
\includegraphics[width=7cm,clip=]{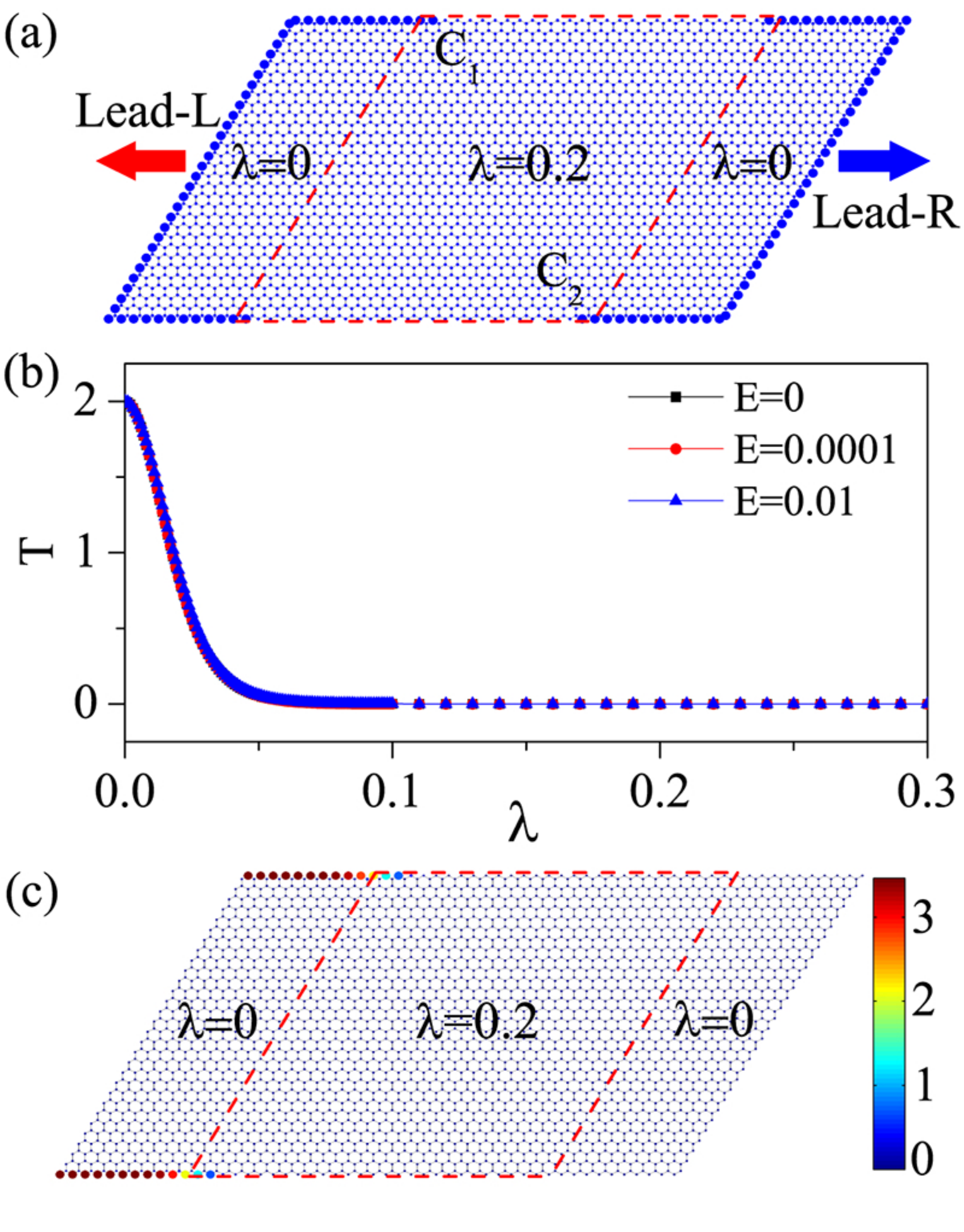}
\caption{(a) Schematic of the setup connected to left and right leads. The system consists of zigzag ribbons without Zeeman fields and the diamond-shaped flake with $\lambda=0.2$ highlighted by red dashed lines. The eigenfunction distribution for this isolated system at zero energy is shown in blue. (b) Transmission with respect to the Zeeman field strength for different energies. (c) Partial LDOS incident from the left lead.}\label{figs1}
\end{figure}

In the setup of Fig.\ref{figs1}(a), the diamond-shaped central flake with nonzero Zeeman field is connected to left and right leads. The lead Hamiltonian follows the conventional Kane-Mele model defined in Eq.(3), and spin-helical edge states propagate in both zigzag leads when the electron energy is inside the bulk gap. In Fig.\ref{figs1}(a), we plot the zero-energy eigenfunction distribution for the isolated system, and blue dots along zigzag boundaries of those $\lambda=0$ regions illustrate the topological edge states. The edge states disappear in the $\lambda=0.2$ region due to the breaking of time-reversal symmetry. Topological corner states are absent in this setup due to two reasons: (1) when the leads are directly coupled to $C_1$ and $C_2$ of the central flake, inside the $\lambda \neq 0$ region there are no corners or boundary intersections where corner states could reside; (2) in the presence of leads, corner states will leak out. We have confirmed that, when a narrow lead of width $W_L=1$ is directly connected to the corner, the corresponding corner state will be destroyed. Besides, the eigenfunction distribution in the $\lambda=0.2$ region is zero, which indicates that electrons can not cross this region. To confirm this point, the transmission as a function of the Zeeman field strength is plotted in Fig.\ref{figs1}(b). It is found that the transmission rapidly decreases from $T=2$ to zero with the increasing of $\lambda$, and $T$ is hardly affected by the variation of electron energies. $T=2$ corresponds to the propagation of two in-gap edge modes, and they are quickly destroyed by the Zeeman field inside the central flake. In Fig.\ref{figs1}(c), we plot the zero-energy partial LDOS for electrons incident from the left lead, which shows only nonzero density of states in the left lead due to the spin-helical edge states.

\subsection*{S2. Setup II: the central region coupled with left and bottom leads}

\begin{figure}
\includegraphics[width=8.5cm ]{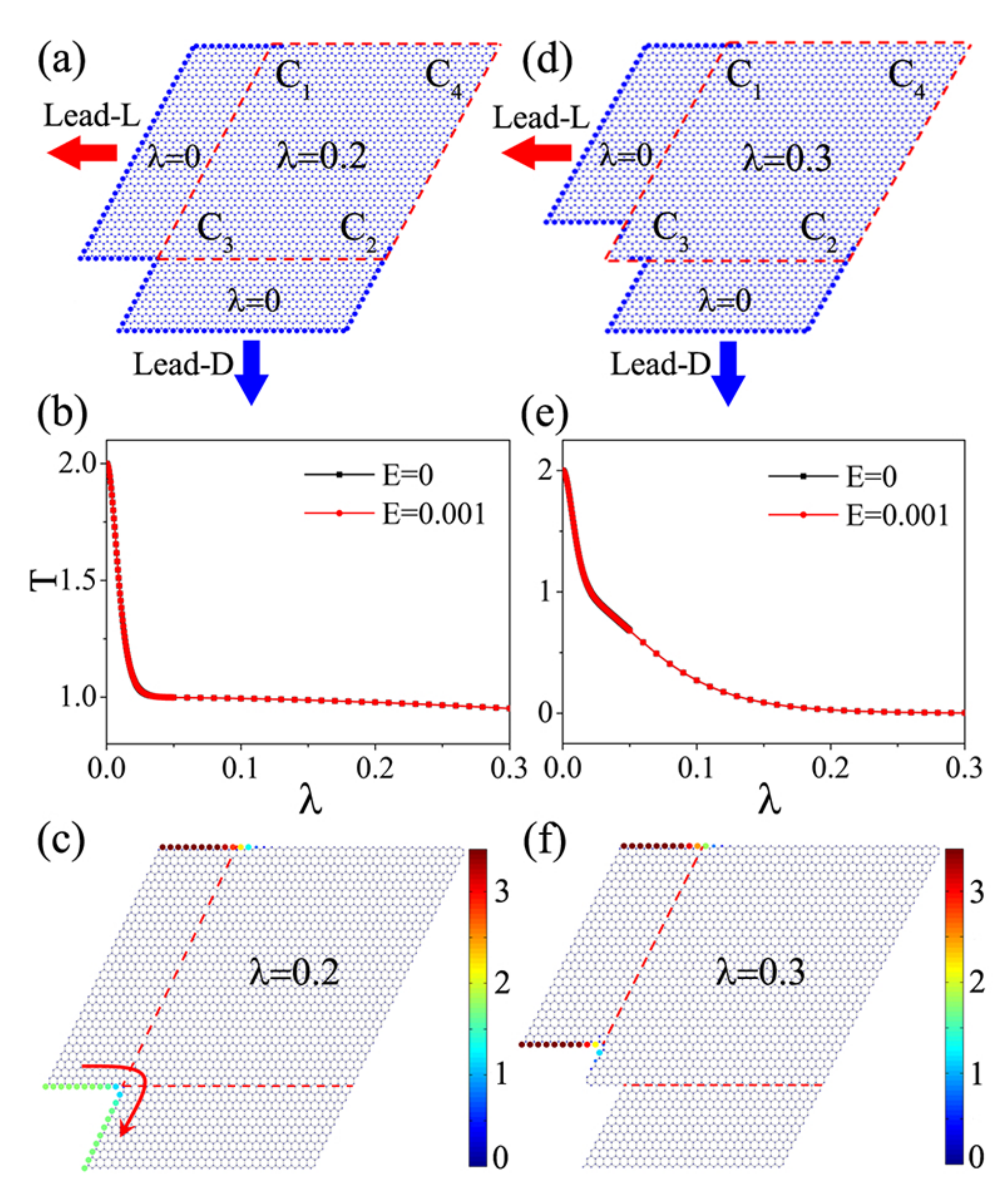}
\caption{ Panels (a) and (d): Two setups coupled with left and bottom leads for lead widths $W_{L}=30$ and $W_L=25$, respectively, accompanied by zero-energy eigenfunction distributions. Panels (b) and (c): Transmission and partial LDOS for the setup in panel (a). Panels (e) and (f): Transmission and partial LDOS for the setup in panel(d).}\label{figs2}
\end{figure}

Two setups are depicted in panels (a) and (d) of Fig.\ref{figs2}, where left and bottom leads are connected to the central region. The zero-energy eigenfunction shows no sign of corner states in Fig.\ref{figs2}(a), which is similar to Setup I. In Fig.\ref{figs2}(b), the transmission of this system rapidly declines from $T=2$ to $T=1$ when the Zeeman field is applied, and slightly decays upon further increasing of $\lambda$. Single conducting channel is found in Fig.\ref{figs2}(c) labeled by the red arrow, which connects the left and down leads near the $C_3$ corner. In the presence of strong Zeeman fields, the $C_3$ corner with an acute angle does not host electronic states, which is responsible for the slow decay of transmission from $T=1$ in this setup. Compared with the single corner-state setup in Fig.2, though no corner state is found at the $C_1$ corner, it can still support certain electronic states and result in a robust $T$=1 plateau against the Zeeman field as shown in Fig.2(b). The single conducting channel in Fig.\ref{figs2}(c) can be easily destroyed by considering narrower leads and exposing the $C_3$ corner in vacuum. Such a setup is presented in Fig.\ref{figs2}(d), where the eigenfunction distribution at the $C_3$ corner is depleted by the in-plane Zeeman field. In Fig.\ref{figs2}(e) and (f), it is found that the transmission of this setup directly drops to zero at large Zeeman fields, which is consistent with the partial LDOS that the propagating channel is closed.

\subsection*{S3. Single Corner-State Setup: features of resonant tunneling}

\begin{figure}
\includegraphics[width=8cm]{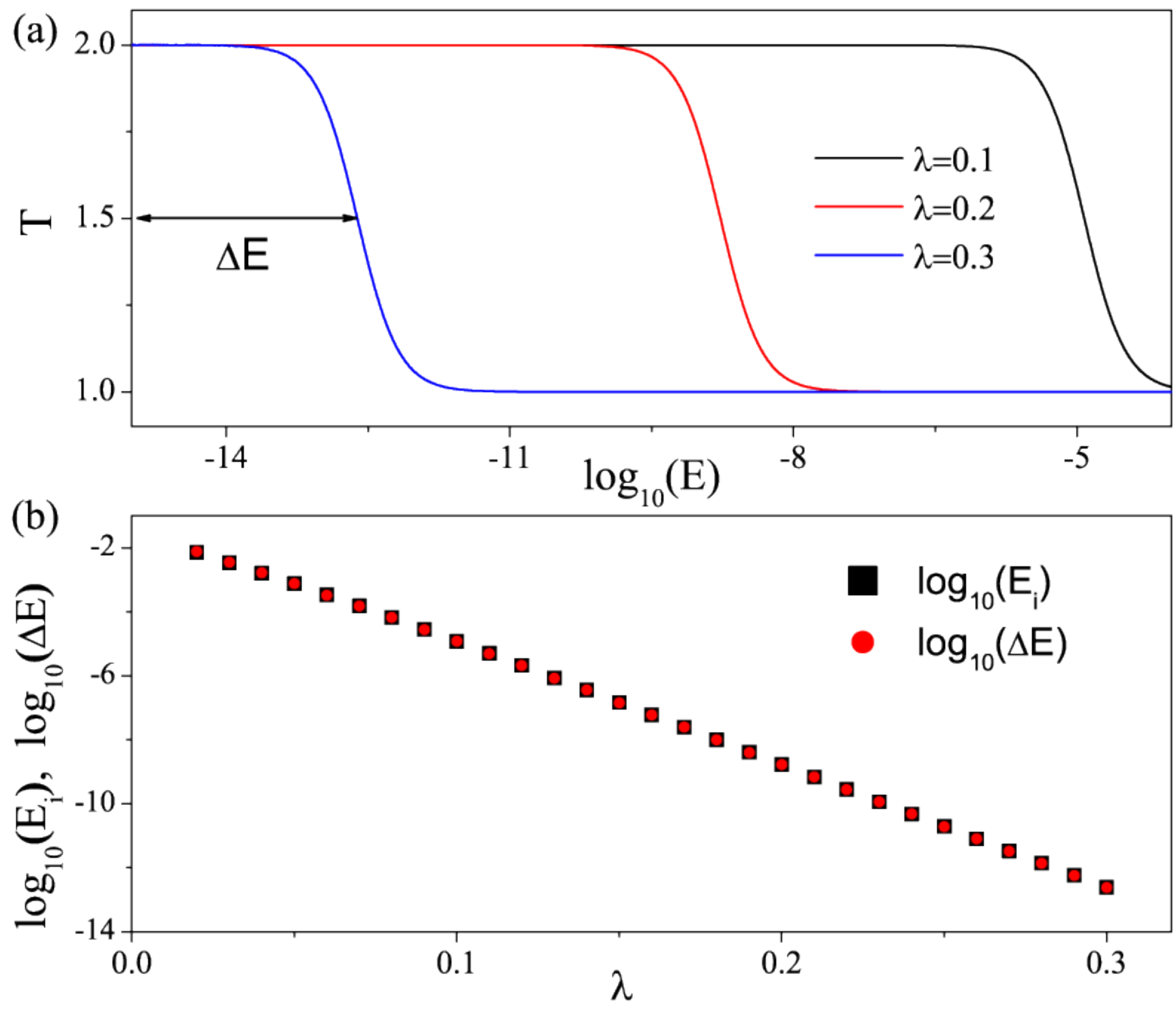}
\caption{(a) Transmission versus the incident energy $E$ for several Zeeman field strengths of the single corner-state setup. The black arrow corresponds to the half-width energy $\Delta E$ at half-maximum $T=1.5$. (b) $\Delta E$ and the imaginary part $E_{i}$ of $\epsilon_{\gamma}$ in Eq.(\ref{resHam}) with respect to $\lambda$ for the $T=2$ resonant peak. }\label{figs3}
\end{figure}
To demonstrate the resonant feature of the $T=2$ transmission plateau for the single corner-state setup shown in Fig.2, we plot the dependence of transmission of this setup on the incident energy in Fig.\ref{figs3}(a). Note that the energy is on the logarithmic scale in order to illustrate the extremely shape peak near $E=0$. The $T=2$ transmission peak is easily destroyed by either changing the incident energy or increasing the Zeeman field strength, showing its resonant nature. Meanwhile, the $T=1$ plateau is robust against both $\lambda$ and $E$, since it originates from the edge state connecting left and top leads at the $C_1$ corner in Fig.2(c). To further confirm the resonant nature, we define $\Delta E$ as the half-width energy at half-maximum of the resonant peak ($T=1.5$). Apparently, the stronger the Zeeman field, the smaller the $\Delta E$. A stronger Zeeman field leads to more localized corner states, and it is easier to close the resonant channel through the $C_2$ corner. It is well known that in resonant tunneling, electrons can tunnel through the system if the incident energy $E$ is in line with the resonant level $E_{res}$. $E_{res}$ can be calculated by solving the eigenvalue problem of an effective Hamiltonian $H_{eff}=H+\Sigma^{r}$ for the open system, which satisfies~\cite{Datta1997}
\begin{equation}
  [H+\Sigma^{r}] \psi_{\gamma} = \epsilon_{\gamma} \psi_{\gamma}. \label{resHam}
\end{equation}
The eigenenergies of this effective Hamiltonian are complex, i.e., $\epsilon_{\gamma}=E_{r}+iE_{i}$. The real part $E_{r}$ corresponds to the resonant level $E_{res}$, which is zero in the zero-energy corner-state-mediated resonant tunneling process. The corresponding imaginary part $E_{i}$ is equivalent to $\Delta E$, the half-width energy at half-maximum of the resonant peak. This well-established theory has been verified in various tunneling systems~\cite{1983Transmission,1984Physics,1985Effect,1992Transmission}. In Fig.\ref{figs3}(b), we plot $\Delta E$ as well as $E_{i}$ against the Zeeman field strength for the $T=2$ peak in Fig.\ref{figs3}(a). $\Delta E$ is measured at the transmission peak and $E_{i}$ is obtained through diagonalizing the effective Hamiltonian $H_{eff}$. Clearly, $\Delta E$ perfectly matches with $E_{i}$ in a wide range of $\lambda$, which further consolidates the resonant tunneling characteristic of the $T=2$ transmission peak.

\subsection*{S4. Double Corner-State Setup: the influence of lead width on transmission}

\begin{figure}
\includegraphics[width=8cm ]{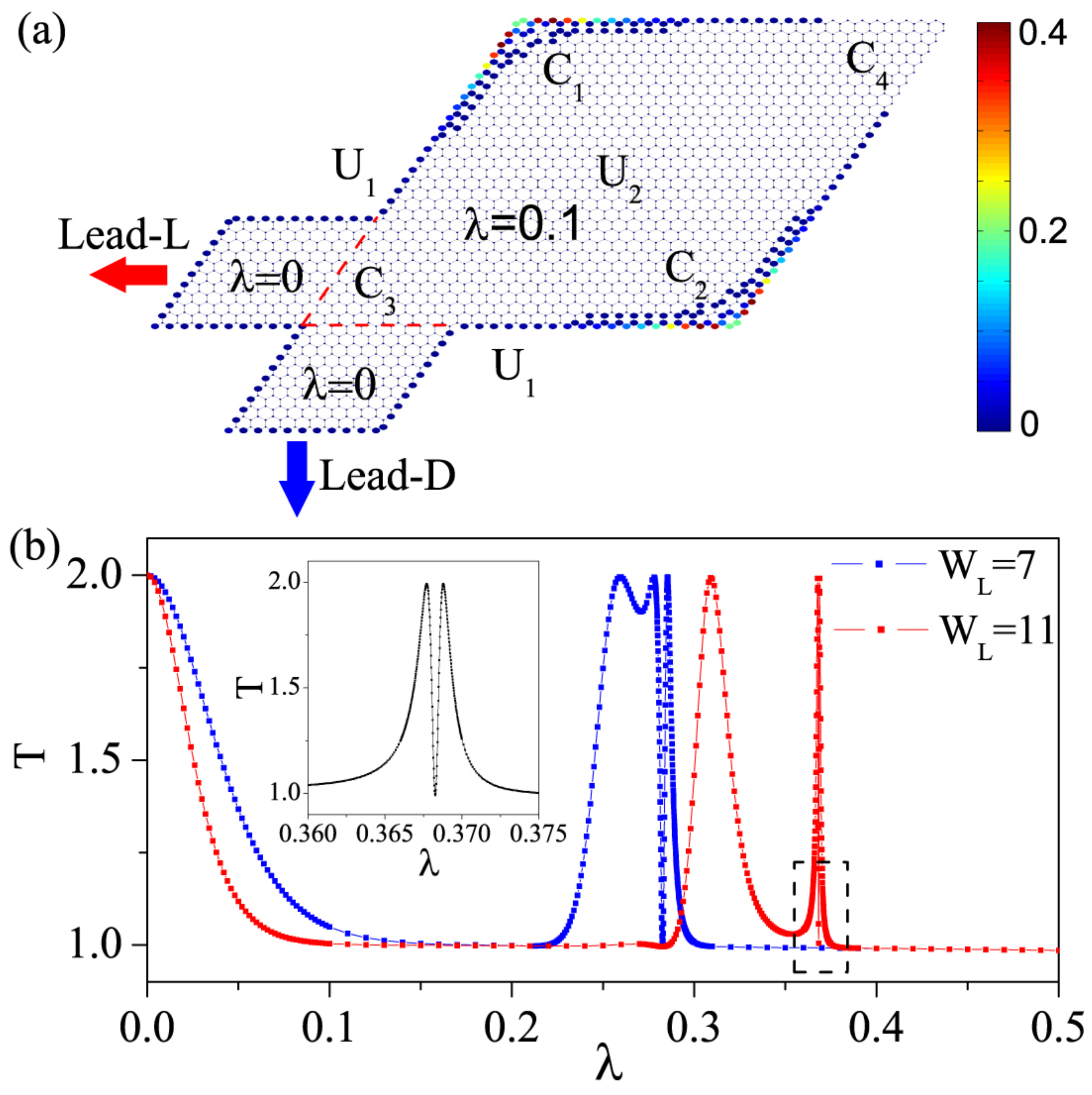}
\caption{(a) The double corner-state setup connected to narrower left and bottom leads around the $C_3$ corner. (b) Transmission for different lead widths $W_L$. The inset highlights the region in the dash rectangle.}\label{figs4}
\end{figure}

For the double corner-state setup shown in Fig.\ref{figs4}(a), narrower left and bottom leads are connected to the diamond-shaped flake around the $C_3$ corner to ensure weak impact on the corner states, which are clearly seen at $C_1$ and $C_2$. We evaluate the influence of lead width on the transmission of this setup. Transmission as a function of the Zeeman field strength for two lead widths are displayed in Fig.\ref{figs4}(b). These two systems share similar features: the transmission drops from $T=2$ in the presence of a Zeeman field, and reaches the $T=1$ plateau with the increasing of $\lambda$, and finally exhibits multiple $T$=2 resonant peaks when the Zeeman field is further enhanced. For a narrower lead of width $W_L=7$, the multiple resonant peaks are wide and close to each other. While for the lead width $W_L=11$, single peak is separated from double peaks shown in the inset. A larger Zeeman field is required to detect the first $T$=2 resonant peak for $W_L=11$. Dynamic details of the double resonant peaks are revealed in the following.

\begin{figure}[tbp]
\includegraphics[width=\columnwidth ]{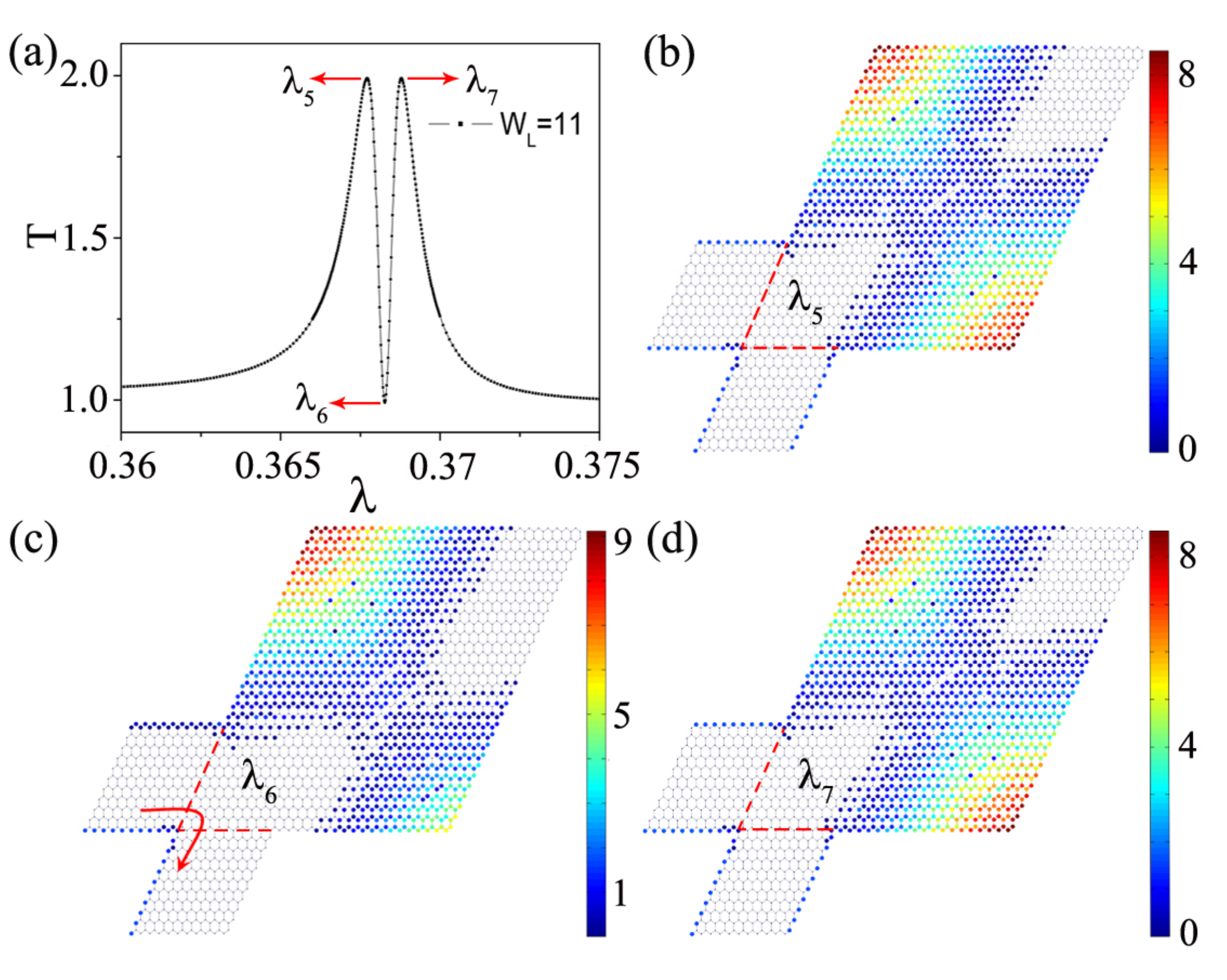}
\caption{(a) The double transmission peaks for lead width $W_L=11$. $\lambda_5$ to $\lambda_7$ correspond to two peaks and one dip in the transmission spectrum. Panels (b)-(d): Partial LDOS for $\lambda_5$ to $\lambda_7$.}\label{figs5}
\end{figure}

The double transmission peaks with respect to the Zeeman field strength is highlighted in Fig.\ref{figs5}(a), where two sharp peaks is separated by a dip. The transmission is $T$=2 for $\lambda_5$ and $\lambda_7$, and $T$ drastically drops to exactly $T$=1 at $\lambda_6$ forming the dip. The partial LDOS in Fig.\ref{figs5}(b) shows that, at $\lambda_5$, a symmetrical dimer state mediates a resonant tunneling channel, which crosses almost the entire bulk of the diamond-shaped flake. In contrast, the precursor of the dimer state has asymmetrical distributions, which is exemplified at $\lambda_6$ where this delicate dimer state is "destroyed" by slight increasing of the Zeeman field strength. In Fig.\ref{figs5}(c), it is obvious that the resonant channel is closed at $\lambda_6$ and single edge-state channel (red arrow) is conducting. Remarkably, the dressing of incoming electrons (albeit asymmetrical) around both corner states is still present, which is preparing for the next resonance. However, the wavefunction becomes asymmetrical, which is typical for the precursor of the dimer state. When the Zeeman field is lightly enhanced to $\lambda_7$, the dimer state emerges again and leads to the second $T=2$ resonant peak. These numerical results vividly demonstrate the dynamic nature of the dimer state induced by electron-dressed corner states. The same dimer state repeatedly emerges in the resonant window when the system is tuned by the Zeeman field, which gives rise to multiple resonant peaks in the transmission spectrum. Finally, we note that the total DOS characterizing the bonding of the dimer state is just the dwell time of incoming electrons, which is proportional to the sharpness of the resonance. A sharper resonance leads to tighter bonding of the dimer state. To summarize, three features of the dimer state are identified: (1) the dressing of incoming electrons bridges two corner states; (2) the scattering wavefunction symmetrically distributes around two corner states; (3) a sharper resonance gives tighter bonding of the dimer state.

\subsection*{S5. Size effect of the diamond-shaped flake}

\begin{figure}
\includegraphics[width=8cm ]{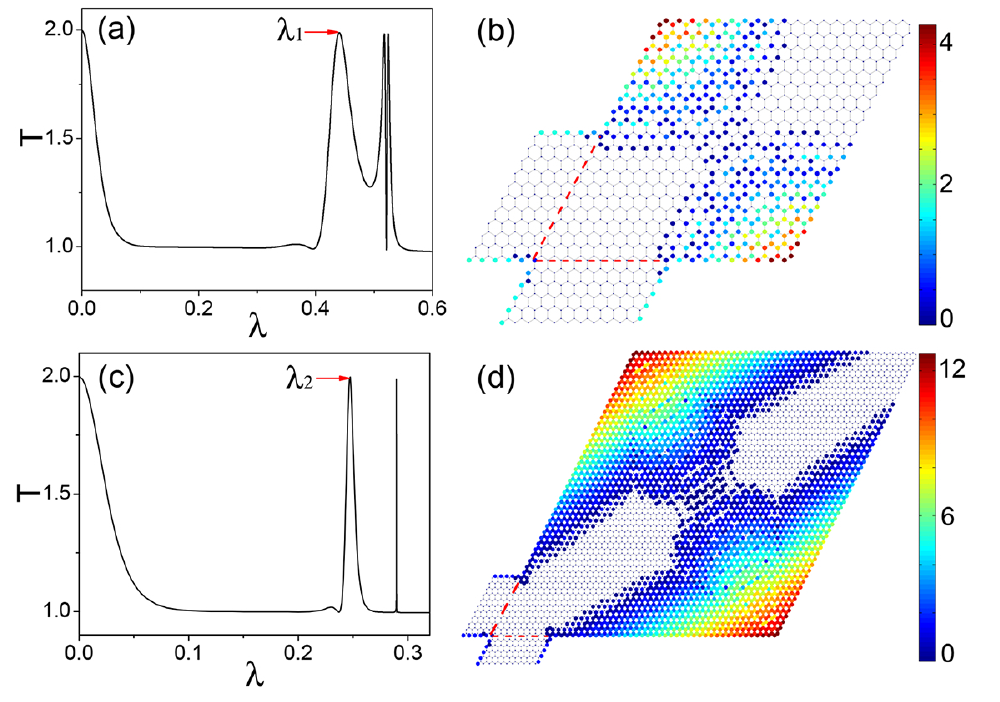}
\caption{(a) Transmission versus the Zeeman field strength for the system size $20a \times 20a$. (b) Partial LDOS corresponding to $\lambda_1$=0.44. (c) Transmission as a function of $\lambda$ for the system size $50a \times 50a$. (d) Partial LDOS corresponding to $\lambda_2$=0.247. The lead width is fixed at $W_L=11$.}\label{figs6}
\end{figure}

To evaluate the size effect of the diamond-shaped flake, we consider two system sizes in the double corner-state setup, which are $20a \times 20a$ and $50a \times 50a$, respectively. The lead width is fixed at $W_L=11$. The corresponding transmission and partial LDOS results are shown in Fig.\ref{figs6}. Combined with numerical results for the system size $30a \times 30a$, we find that all systems have multiple resonant peaks at large Zeeman field strengths, which corresponds to the dimer state induced by electron-dressed corner states. With the increasing of system size, the resonant peaks become narrower. At the same time, a relatively smaller Zeeman field strength is required to observe the first resonant peak for a larger system. Therefore, the dimer-state-mediated resonant tunneling is a common feature in the double corner-state setup, regardless of the system size.

\end{document}